\global\def\draftcontrol{0}

   \def\versionno{ superpotential }

\catcode`\@=11

\newcommand\makepapertitle{\par

  \begingroup
    \renewcommand\thefootnote{\@fnsymbol\c@footnote}%
 \newpage
     \global\@topnum\z@   
     \@makepapertitle
     \thispagestyle{empty}\@thanks
  \endgroup
  \setcounter{footnote}{0}%
  \global\let\thanks\relax
  \global\let\makepapertitle\relax
  \global\let\@makepapertitle\relax
  \global\let\@thanks\@empty
  \global\let\@author\@empty
  \global\let\@date\@empty
  \global\let\@title\@empty
  \global\let\title\relax
  \global\let\author\relax
  \global\let\date\relax
  \global\let\and\relax
  \def\version{\let\version\@version\@gobble}
}
\def\@makepapertitle{%
  \newpage
   \ifnum\draftcontrol=1 {}
   \version\versionno
   \vskip 5.5em%
   \else
   \hfill\hbox to 3.5cm {\parbox{5cm}{\@pubnum}\hss}%
   \vskip 6.5em%
   \fi
   \begin{center}%
   \let \footnote \thanks
      {\hskip -0\textwidth \hbox to 1\textwidth%
        {\centerline{\Large\bf{\noindent%
    \parbox[t]{1.3\textwidth}{\begin{center}\@title\end{center}}}}}}%
     \vskip 1.5em%
     {\normalsize
       \lineskip .5em%
       \begin{tabular}[t]{c}%
         \@author
       \end{tabular}\par}%
     \vskip 1.5em%
     {\@bstract}%
     \end{center}%
     \vfill
     \@date%
     \vskip 1.5em%
   \par
}

\gdef\@pubnum{}
\def\pubnum#1{%
  \gdef\@pubnum{#1}}

\gdef\@bstract{}
\def\Abstract#1{%
  \gdef\@bstract{%
   \parbox{\textwidth-0pc}{%
   \centerline{\bf Abstract}\penalty1000
   \noindent
   \renewcommand\baselinestretch{1.0}
   {#1}}}
}

\gdef\@email{}
\def\email#1{%
   \gdef\@email{%
   Email: {\tt #1}}
}

\def\ps@paper{\let\@mkboth\@gobbletwo%
     \ifnum\draftcontrol=1
        \def\@oddfoot{\hbox to \textwidth{\tiny \versionno \hfil\tiny\draftdate}%
        \hskip -\textwidth \hbox to \textwidth{\hfil\rm\thepage\hfil}}%
     \else\def\@oddfoot{\hbox to \textwidth{\hfil\rm\thepage\hfil}}
     \fi
     \let\@evenfoot\@oddfoot
}

\def\body{\clearpage
          \pagestyle{paper}
        }

\def\@version#1{\ifnum\draftcontrol=1
\typeout{}\typeout{#1}\typeout{}
\vskip3mm\centerline{\hbox{\fbox{\normalsize{\tt DRAFT -- #1 -- }
                   {\draftdate}}}}\vskip3mm
\fi}
\let\version\@version
\long\def\eqlabel#1{\ifnum\draftcontrol=1
                    \tag@false  
                    \tag*{(\theequation) \hbox to -0.2cm{\hspace{0cm}\small{#1}\hss}}
                    \refstepcounter{equation}
                    \edef\@currentlabel{\theequation}
                    \ltx@label{#1}
                    \else
                    \label{#1}
                    \fi
                    }
\let\st@bibitem\@bibitem
\let\st@lbibitem\@lbibitem
\ifnum\draftcontrol=1
  \def\@bibitem#1{%
    \st@bibitem{#1}\a@@label{#1}\ignorespaces}
  \def\@lbibitem[#1]#2{%
    \st@lbibitem[#1]{#2}\a@@label{#2}\ignorespaces}
  \def\a@@label#1{%
    \gdef\a@lab{\smash{\normalfont\small#1}}
    \ifvmode
      \if@inlabel
        \global\setbox\@labels\hbox{%
          \llap{\a@lab\let\a@lab\relax
                \kern\@totalleftmargin\kern\marginparsep}%
          \box\@labels}%
      \fi
    \fi}
\fi

\documentclass[12pt,letterpaper]{article}
\usepackage{amsmath}
\usepackage{amsmath}
\usepackage{amsmath}
\usepackage{amssymb}
\usepackage{amssymb}
\usepackage{amssymb}
\usepackage{amsmath,bm,amsfonts,amssymb,array,calc,amsthm,rotating,cite}
\usepackage{epsfig,psfrag}
\usepackage{graphicx}
\usepackage{color}
\usepackage[colorlinks=true]{hyperref}
\usepackage[all]{xy}

\renewcommand\baselinestretch{1.25}
\renewcommand\section{\@startsection {section}{1}{\z@}%
                                   {-3.5ex \@plus -1ex \@minus -.2ex}%
                                   {2.3ex \@plus.2ex}%
                                   {\normalfont\large\bfseries}}
\renewcommand\subsection{\@startsection{subsection}{2}{\z@}%
                                   {-3.25ex\@plus -1ex \@minus -.2ex}%
                                   {1.5ex \@plus .2ex}%
                                   {\normalfont\normalsize\bfseries}}
\renewcommand\subsubsection{\@startsection{subsubsection}{3}{\z@}%
                                   {-3.25ex\@plus -1ex \@minus -.2ex}%
                                   {1.5ex \@plus .2ex}%
                                   {\normalfont\normalsize\it}}
\renewcommand\paragraph{\@startsection{paragraph}{4}{\z@}%
                                   {-3.25ex\@plus -1ex \@minus -.2ex}%
                                   {1.5ex \@plus .2ex}%
                                   {\normalfont\normalsize\bf}}
\renewcommand\subparagraph{\@startsection{subparagraph}{5}{\z@}%
                                   {-1.25ex\@plus -1ex \@minus -.2ex}%
                                   {0ex \@plus .2ex}%
                                   {\normalfont\normalsize\it}}

\numberwithin{equation}{section}

\renewcommand*\l@section[2]{%
  \ifnum \c@tocdepth >\z@
    \addpenalty\@secpenalty
    \addvspace{.5em \@plus\p@}%
    \setlength\@tempdima{1.5em}%
    \begingroup
      \parindent \z@ \rightskip \@pnumwidth
      \parfillskip -\@pnumwidth
      \leavevmode \bfseries
      \advance\leftskip\@tempdima
      \hskip -\leftskip
      #1\nobreak\hfil \nobreak\hb@xt@\@pnumwidth{\hss #2}\par
    \endgroup
  \fi}
\renewcommand*\l@subsection{\addvspace{.0em \@plus\p@}\@dottedtocline{2}{1.5em}{2.3em}}
\renewcommand*\l@subsubsection{\addvspace{-.2em \@plus\p@}\@dottedtocline{3}{3.8em}{3.2em}}

\definecolor{refcol}{rgb}{0.0,0.0,0.2}
\definecolor{eqcol}{rgb}{.2,0,0}
\definecolor{purple}{cmyk}{0,1,0,0}

\gdef\@citecolor{refcol} \gdef\@linkcolor{eqcol}
\gdef\@urlcolor{refcol}
\def\colorlinkspurple{\gdef\@urlcolor{purple}}
\def\colorlinksblue{\gdef\@urlcolor{blue}}
\def\colorlinksred{\gdef\@urlcolor{red}}

\def\revise#1       {\raisebox{-0em}{\rule{3pt}{1em}}%
                     \marginpar{\raisebox{.5em}{\vrule width3pt\
                     \vrule width0pt height 0pt depth0.5em
                     \hbox to 0cm{\hspace{0cm}{%
                     \parbox[t]{4em}{\raggedright\footnotesize{#1}}}\hss}}}}

\catcode`\@=12
\newcommand{\bqa}{\begin{eqnarray}}
\newcommand{\eqa}{\end{eqnarray}}
\begin{document}

\title{
 Ooguri-Vafa Invariants and Off-shell Superpotentials\\ 
 of Type II/F-theory compactification}

\author{
Feng-Jun Xu,~~~Fu-Zhong Yang\footnote{Corresponding author~~~
E-mail: fzyang@gucas.ac.cn} \\[0.2cm]
\it College of Physical Sciences, Graduate University of Chinese Academy of Sciences\\
\it   YuQuan Road 19A, Beijing 100049, China}

\Abstract{~~~~In this paper, we make a further step of \cite{Feng}
and calculate off-shell superpotential of two Calabi-Yau manifold
with three parameters by integrating the period of subsystem. We
also obtain the Ooguri-Vafa invariants with open mirror symmetry. }

\makepapertitle

\body

\version\versionno

\vskip 1em

\newpage

\section{Introduction}

~~~~When Type II string theory compactfying on Calabi-Yau threefold
with D-brane and background flux, the superpotential will be
generated which in general can divided two parts--one originated
from D-brane and the other from flux. The superpotential also plays
an important role in mathematics which generates the Ooguri-Vafa
invariants and counting the number of disks and sphere instantons.

 For D5-brane wrapped the whole Calabi-Yau
threefold, the holomorphic Chern-Simons theory \cite{Witten:1992fb}
\bqa
  \mathcal{W}=\int_{X}\Omega^{3,0}\wedge \text{Tr}[A\wedge \bar{\partial}A+\frac{2}{3}A\wedge A\wedge A]
 \eqa
gives the brane superpotential $\mathcal{W}_{brane}$, where $A$ is
the gauge field with gauge group $U(N)$ for $N$ D5-branes. When
reduced dimensionally, the low-dimenaional brane superpotentials can
be obtained as \cite{Lerche:2003hs,Aganagic:2000gs}
 \bqa
\mathcal{W}_{brane}
=N_{\nu}\int_{\Gamma^{\nu}}\Omega^{3,0}(z,\hat{z})=\sum_{\nu}N_{\nu}\Pi^{\nu}
 \eqa
 where $\Gamma^{\nu}$ is a special Lagrangian 3-chain and
$(z,\hat{z})$ are closed-string complex structure moduli and D-brane
moduli from open-string sector, respectively.

    The background fluxes $H^{(3)}=H_{RR}^{(3)}+\tau H_{NS}^{(3)}$, which take values in the integer cohomology group $H^3(X,\mathbb{Z})$, also break the supersymmetry $N=2$ to $N=1$.
 The $\tau=C^{(0)}+ie^{-\varphi}$ is the complexified Type IIB coupling field. Its
 contribution to superpotentials is \cite{Mayr:2000hh,Taylor:1999ii}
 \bqa
 \mathcal{W}_{flux}(z)\ =\int_{X}
 H_{RR}^{(3)}\wedge  \Omega^{3,0} = \sum_\alpha N_\alpha\cdot\Pi^\alpha(z)\ ,
 \qquad N_\alpha \in Z.
\eqa

  The contributions of D-brane and background flux (here the NS-flux ignored) give together the general
  form of superpotential as follow\cite{Lerche:2002ck,Lerche:2002yw}
  \bqa
\mathcal{W}(z,\hat{z})=\mathcal{W}_{brane}(z,\hat{z})+\mathcal{W}_{flux}(z)=\sum_{\gamma_{i}\in
H^3(Z^\ast,\mathcal{H})}N_{i}\Pi_{i}(z,\hat{z}) \eqa where
$N_{i}=n_{i}+\tau m_{\sigma}$, $\tau$ is the dilaton of type II
string and $\Pi_{i} $ is a relative periods defined in a relative
cycle $\Gamma\in H_3(X,D)$ whose boundary is wrapped by D-branes and
D is a holomorphic divisor of the Calabi-Yau space. In fact, the
two-cycles wrapped by the D-branes are holomorphic cycles only if
the moduli are at the critical points of the superpotentials. Thus,
the two-cycles are generically not holomorphic. However, according
to the arguments of
\cite{Lerche:2002ck,Lerche:2002yw,Jockers:2008pe}, the
non-holomorphic two-cycles can be replaced by a holomorphic divisor
D of the ambient Calabi-Yau space with the divisor D encompassing
the two-cycles.

Geometrically speaking, when varying the complex structure of
Calabi-Yau space, a generic holomorphic curve will not be
holomorphic with the respect to the new complex structure, and
becomes obstructed to the deformation of the bulk moduli. The
requirement for the holomorphy gives rise to a relation between the
closed and open string moduli. Physically speaking, it turns out
that the obstruction generates a superpotential for the effective
theory depending on the closed and open string moduli.

  The off-shell
tension of D-branes, $\mathcal{T}(z,\hat{z})$, is equal to the
relative period \cite{Witten:1997ep,Lerche:2002ck,Lerche:2002yw}
\bqa \Pi_{\Sigma}=\int_{\Gamma_{\Sigma}}\Omega(z,\hat{z}) \eqa
 which
measures the difference between the value of on-shell
superpotentials for the two D-brane configurations\bqa
\mathcal{T}(z,\hat{z})=\mathcal{W}(C^+)-\mathcal{W}(C^-) \eqa  with
$\partial \Gamma_{\Sigma}=C^+-C^-$. The domain wall tension is
\cite{Alim:2010za} \bqa
T(z)=\mathcal{T}(z,\hat{z})\mid_{\hat{z}=\text{critic points}}\eqa
where the critic points correspond to $\frac{dW}{d\hat{z}}=0$
\cite{Witten:1997ep} and the $C^{\pm}$ is the holomorphic curves at
those critical points. The critical points are alternatively defined
as the Nother-Lefshetz locus \cite{Clemens}
 \bqa
\mathcal{N}=\{(z,\hat{z})\mid\pi(z,\hat{z};\partial\Gamma(z,\hat{z}))\equiv
0\}
 \eqa
where \bqa \eqlabel{relative}
\pi(z,\hat{z};\partial\Gamma(z,\hat{z}))=\int_{\partial\Gamma}\omega_{\hat{a}}^{(2,0)}(z,\hat{z}),~~~\hat{a}=1,...,\text{dim}(H^{2,0}(D))
\eqa and $\omega_{\hat{a}}^{(2,0)}$ is an element of the cohomology
group $H^{(2,0)}(D)$. At those critical points, the domain wall
tensions are also known as normal function giving the Abel-Jacobi
invariants \cite{Clemens,
Alim:2010za,Li:2009dz,Morrison:2007bm,Griffiths}

  The Superpotential can be calculated by study the Hodge variation on
  the cohomology group $\Gamma\in H_3(X,D)$. The flat Gauss-Manin
  connection on this cohomology group can determine the mirror map
  between A-model and B-model. By the mirror symmetry, we can also
  obtain the Ooguri-Vafa invariants.

  The purpose of this note is to calculate the off-shell
  superpotential which at the critical point equal to the
  domain wall tensions (on-shell superpotential) that have been obtained in the previous work
  \cite{Feng}.

\section{Generalized GKZ system and Differential Operators}

The period integrals can be written as
   \bqa
   \Pi_i=\int_{\gamma_i}\frac{1}{P(a,X)}{\prod_{j=1}^n \frac{d
   X_j}{X_j}}.
   \eqa
   where $P$ is the hypersurface equation and $a_i$ is the moduli determining the complex structure in B-model. See more in \cite{Feng}.
   According to the refs.\cite{Batyrev:1993wa,Batyrev:1994hm}, the
   period integrals can be annihilated by differential operators
   \bqa
   \eqlabel{operator}
   \begin{split}
   &\mathcal{L}(l)=\prod_{l_i>0}(\partial_{a_i})^{l_i}-\prod_{l_i<0}(\partial_{a_i})^{l_i}\\
   &\mathcal{Z}_k=\sum_{i=0}^{p-1}\nu_{i,k}^{\ast}\vartheta_i,
   \qquad \mathcal{Z}_0=\sum_{i=0}^{p-1}\vartheta_i-1
   \end{split}
   \eqa
where $\vartheta_i=a_i\partial_{a_i}$. As noted in
refs.\cite{Hosono:1993qy}\cite{Alim:2009bx}, the equations
$\mathcal{Z}_k\Pi(a_i)=0$ reflex the invariance under the torus
action, defining torus invariant algebraic coordinates $z_a$ on the
moduli space of complex structure of $X$ \cite{Alim:2010za}: \bqa
\eqlabel{coordinates} z_a=(-1)^{l_0^{a}}\prod_{i}a_i^{l_i^{a}}\eqa
where $l_a$, ~$a=1,...,h^{2,1}(X)$ is generators of the Mori cone,
one can rewrite the differential operators $\mathcal{L}(l)$ as
\cite{Batyrev:1994hm,Hosono:1993qy,Alim:2010za}
    \bqa
    \mathcal{L}(l)=\prod_{k=1}^{l_0}(\vartheta_0-k)\prod_{l_i>0}\prod_{k=0}^{l_i-1}(\vartheta_i-k)-(-1)^{l_0}z_a\prod_{k=1}^{-l_0}(\vartheta_0-k)\prod_{l_i<0}\prod_{k=0}^{l_i-1}(\vartheta_i-k).
    \eqa

   The solution to the GKZ system can be written as \cite{Batyrev:1994hm,Hosono:1993qy,Alim:2010za} \bqa
B_{l^a}(z^a;\rho)=\sum_{n_1,...,n_N\in Z_0^{+}}\frac{\Gamma(1-\sum_a
l_0^a(n_a+\rho_a))}{\prod_{i>0}\Gamma(1+\sum_a
l_i^a(n_a+\rho_a))}\prod_a z_a^{n_a+\rho_a}.\eqa

 In this paper we consider the family of divisors $\mathcal{D}$ with a single open
 deformation moduli $\hat{z}$
  \bqa
    x_1^{b_1}+\hat{z}x_2^{b_2}=0
  \eqa where $b_1,~b_2$ are some appropriate integers. The relative 3-form $\underline{\Omega}:=(\Omega_{X}^{3,0},0)$ and the relative
periods satisfy a set of differential equations
\cite{Lerche:2002ck,Lerche:2002yw,Alim:2009bx,Alim:2010za,Jockers:2008pe}
 \bqa
\mathcal{L}_a(\theta,\hat{\theta})\underline{\Omega}=d\underline{\omega}^{(2,0)}~\Rightarrow~\mathcal{L}_a(\theta,\hat{\theta})\mathcal{T}(z,\hat{z})=0.
\eqa with some corresponding two-form $\underline{\omega}^{(2,0)}$.
The differential operators $\mathcal{L}_a(\theta,\hat{\theta})$ can
be expressed as \cite{Alim:2010za} \bqa
\mathcal{L}_a(\theta,\hat{\theta}):=\mathcal{L}_a^{b}-\mathcal{L}_a^{bd}\hat{\theta}
\eqa for $\mathcal{L}_a^{b}$ acting only on bulk part from closed
sector, $\mathcal{L}_a^{bd}$ on boundary part from open-closed
sector and $\hat{\theta}=\hat{z}\partial_{\hat{z}}$. The explicit
form of these operators will be given in following model. From the
\eqref{relative} one can obtain \bqa \eqlabel{sub} 2\pi
i\hat{\theta}\mathcal{T}(z,\hat{z})=\pi(z,\hat{z}) \eqa for only the
family of divisors $\mathcal{D}$ depending on the $\hat{z}$. Hence
the off-shell superpotential can be obtained by integrating the
period on subsystem $\pi(z,\hat{z})$.

\section{Superpotentials of Hypersurface $X_{24}(1,1,2,8,12)$}
 ~~~~The $X_{24}(1,1,2,8,12)$ is defined as the zero locus of polynomial $P$
 \bqa
P=x_1^{24}+x_2^{24}+x_3^{12}+x_4^{3}+x_5^{2}+\psi
x_1x_2x_3x_4x_5+\phi x_1^6x_2^6x_3^6+\chi x_1^{12}x_2^{12} \eqa

    The GLSM charge vectors $l_a$ are the generators of the Mori cone as follows \cite{Hosono:1993qy}
\begin{equation}
\begin{tabular}{c|c c c c c c c c}
~  & $0$ & $1$& $2$& $3 $ & $4$ & $5$ & $6$ & $7$ \\\hline
$l_1$ & $-6$ & $0$ & $0$& $0$ & $2$ & $3$ & $0$ & $1$ \\
$l_2$ & $0$ & $1$& $1$  & $0$& $0$ & $0$ & $-2$ & $0$ \\
$l_3$ & $0$ &$0$& $0$ & $1$ & $0$ & $0$ & $1$ & $-2$.
\end{tabular}
\end{equation}

    The mirror manifolds can be constructed as an orbifold by the
Greene-Plesser orbifold group acting as $x_i\rightarrow
\lambda_k^{g_{k,i}}x_i$ with weights \bqa \mathbb{Z}_6:
~g_1=(1,-1,0,0,0),~~~\mathbb{Z}_6:~g_2=(1,0,-1,0,0),~~~\mathbb{Z}_3:~g_3=(1,0,0,-1,0)\eqa
where we denotes $\lambda_{1,2}^6=1 ~\text{and} ~\lambda_{3}^3=1 $.

  By the generalized GKZ system, the period on the K3 surface has the
form \bqa
\pi=\frac{c}{2}B_{\{\hat{l}_1,\hat{l}_2,\hat{l}_3\}}(u_1,u_2,u_3;\frac{1}{2},\frac{1}{2},0)=-\frac{4c}{\pi^{\frac{3}{2}}}\sqrt{u_1u_2}u_3+\mathcal{O}((u_1u_2)^{3/2})
\eqa which vanishes at the critical locus $u_2=0$. According to
\eqref{sub}, the off-shell superpotentials can be obtained by
integrating the $\pi$:
  \bqa
  \mathcal{T}_a^{\pm}(z_1,z_2,z_3)=\frac{1}{2\pi
  i}\int\pi(\hat{z})\frac{d\hat{z}}{\hat{z}},
\eqa with the appropriate integral constants\cite{Alim:2010za},
 the superpotentials can be chosen as  $\mathcal{W}^+=-\mathcal{W}^-$. In this convention, the off-shell
 superpotentials can be obtained as
\bqa 2\mathcal{W}^{+}=\frac{1}{2\pi
 i}\int_{-\hat{z}}^{\hat{z}}\pi(\zeta)\frac{d\zeta}{\zeta},~~~W^{\pm}(z_1,z_2,z_3)=\mathcal{W}^{\pm}(z_1,z_2,z_3)|_{\hat{z}=1}
\eqa
  Eventually, The superpotential are
\bqa
\begin{split}
&\mathcal{W}^{\pm}(z_1,z_2,z_3,\hat{z})=\sum_{n_1,n_2,n_3}\frac{\mp cz_1^{\frac{1}{2}+n_1}z_2^{\frac{1}{2}+n_2}z_3^{n_3}\hat{z}^{\frac{-1-2n_2}{2}}\Gamma(6n_1+4)}{\Gamma(2+2n_2)\Gamma(2+2n_1)\Gamma(\frac{5}{2}+3n_1)\Gamma(1+n_3)\Gamma(n_3-2n_2)\Gamma(n_1-2n_3+\frac{3}{2})}\\
&\frac{\{(1-2n_2)
{}_2F_1(-\frac{1}{2}-n_2,-2n_2,\frac{1}{2}-n_2;\hat{z})+\hat{z}(1+2n_2){}_2F_1((\frac{1}{2}-n_2,-2n_2,\frac{3}{2}-n_2;\hat{z}))\}}{4\pi(-1+4n_2^2)}
\end{split}
\eqa

For calculation of instanton corrections, one need to know mirror
  map. The fundamental period $\omega_0$ is solution of the Picard-Fuchs equation which we
  listed in \cite{Feng}.  The flat coordinates in
 A-model at large radius regime are related to the flat coordinates of B-model at large complex structure regime by mirror map $t_i=\frac{\omega_i}{\omega_0},~\omega_i:= D_i^{(1)}
 \omega_0(z,\rho)|_{\rho=0}$. The open-string mirror map are
 \bqa
\begin{split}
 &q_1=z_1+312z_1^2+107604z_1^3-z_1z_3-192z_1^2z_3-z_1z_3^2+\mathcal{O}(z^4)\\
 &q_2=z_2+2z_2^2+5z_2^3+z_2z_4+3z_2^2z_4+z_2^2z_4^2+\mathcal{O}(z^4)\\
 &q_3=z_3+2z_3^2+3z_3^3+120z_3z_1+41580z_1^2z_3+\mathcal{O}(z^4)\\
 &q_4=z_4-z_4^2+z_4^3+\mathcal{O}(z^4)
\end{split}
 \eqa
 here $q_i=e^{2\pi i
t_i}$ and we can obtain the inverse mirror map
 \bqa
\begin{split}
&z_1=q_1-312q_1^2+87084q_1^3+q_1q_3-864q_1^2q_3+q_1q_2q_3+\mathcal{O}(q^4)\\
&z_2=q_2-2q_2^2+3q_2^3+\mathcal{O}(q^4)\\
&z_3=q_3-2q_3^2+3q_3^3-120q_1q_3+10260q_1^2q_3+q_2q_3-120q_1q_2q_3+600q_1q_3^2-4q_2q_3^2+\mathcal{O}(q^4)\\
&z_4=q_4+q_4^2+q_4^3+\mathcal{O}(q^4)
\end{split}
 \eqa

 \begin{table}[!h]
\def\temptablewidth{1.0\textwidth}
\begin{center}
\begin{tabular*}{\temptablewidth}{@{\extracolsep{\fill}}c|ccccc}
$d_4=0,d_3=1$&&&&&\\
$d_1/2\backslash d_2/2$ & 1 &3     &5     &7    &9 \\\hline
 1              &  1          &  0                  & 0             &0                  &$\frac{-5}{2}$    \\
  3             & -848            &0                 & 0     & 0               & 2120  \\
   5            &  -270978         & 0             & 0     &0           & 677445   \\
   7           & -4107040      & 0      & 0  &0      &10267600   \\
   9            & -4859101222  &0 &0 &0 &12147753055
       \end{tabular*}
       {\rule{\temptablewidth}{1pt}}
\end{center}
       \end{table}
\begin{table}[!h]
\def\temptablewidth{1.0\textwidth}
\begin{center}
\begin{tabular*}{\temptablewidth}{@{\extracolsep{\fill}}c|ccccc}
$d_4=0,d_3=2$&&&&&\\
$d_1/2\backslash d_2$ & 1        &3         &5    &7    &9
\\\hline
 1                &$\frac{-9 }{16}$      & $\frac{-9 }{16}$      & 0        &0        &$\frac{45 }{32}$  \\
  3              & $\frac{521}{2}$         &$\frac{521}{2}$         & 0         & 0      & $\frac{-2605}{4}$  \\
   5              & $\frac{-1397265}{8}$     &$\frac{-2506065}{8}$& 167400     &-195120 & $\frac{7890645}{16}$   \\
   7              &100877911                  & 205105111        & -118540800 &142047360  &$\frac{-553418675}{2}$   \\
   9              & $\frac{-160323502433}{8}$ &$\frac{226729748767}{8}$ &-64409331600  &71920841760 &$\frac{251804856805}{16}$
       \end{tabular*}
       {\rule{\temptablewidth}{1pt}}
\end{center}
       \end{table}

\begin{table}[!h]
\def\temptablewidth{1.0\textwidth}
\begin{center}
\begin{tabular*}{\temptablewidth}{@{\extracolsep{\fill}}c|ccccc}
$d_4=1,d_3=1$&&&&&\\
$d_1/2\backslash d_2$ & 1        &3         &5    &7    &9
\\\hline
 1                &$\frac{-29 }{18}$   & 0      & 0       &$\frac{-7 }{2}$      &$\frac{-35 }{36}$  \\
  3              & $\frac{12296}{9}$      & 0      &0     & 2968      & $\frac{7420}{9}$  \\
   5              & $\frac{1309727}{3}$   & 0     &5130    &943293 & $\frac{1611485}{6}$   \\
   7              & $\frac{59552080}{9}$   & 0    & -3734640     &18109280 &$\frac{2324840}{9}$   \\
   9              & $\frac{70456967719}{9}$& 0    &-1890907740   &18897762017 &$\frac{50997932065}{18}$
       \end{tabular*}
       {\rule{\temptablewidth}{1pt}}
       \tabcolsep 0pt \caption{Disc invariants $n_{d_1,d_2,d_3,d_4}$ for the off-shell
superpotential $W_1$ of the 3-fold $\mathbb{P}_{1,1,2,8,12}[24]$}.
\vspace*{-12pt}
\end{center}
       \end{table}

  Using the modified multi-cover formula\cite{Aganagic:2000gs} for
this case
  \bqa
\frac{\mathcal{W}^{\pm}(z(q))}{w_0(z(q))}=\frac{1}{(2
i\pi)^2}\sum_{k~
 odd}~\sum_{d_3 ,d_4, d_{1,2} odd\geq
0}n_{d_1,d_2,d_3,d_4}^{\pm}\frac{q_1^{kd_1/2}q_2^{kd_2/2}q_3^{kd_3}q_4^{kd_4}}{k^2},
\eqa

the superpotentials $\mathcal{W}^+$ give Ooguri-Vafa invariants
$n_{d_1,d_2,d_3,d_4}$ for the normalization constants
  $c=1$, which are listed in Table.1.

\section{Superpotential of Hypersurface $X_{12}(1,1,1,3,6)$}
~~~~The $X_{12}(1,1,1,3,6)$ is defined as zero locus of $P$: \bqa
\begin{split}
P=x_1^{12}+x_2^{12}+x_3^{12}+x_4^4+x_5^{2}+\psi x_1x_2x_3x_4x_5+\phi
x_1^4x_2^4x_3^4
\end{split}
\eqa The GLSM charge vectors in this case are \cite{Hosono:1993qy}
\begin{equation}
\begin{tabular}{c|c c c c c c c}
~  & $0$ & $1$& $2 $ & $3$ & $4$ & $5$ & $6$ \\\hline
$l_1$ & $-4$ & $0$ & $0$ & $0$ & $1$ & $2$ & $1$ \\
$l_2$ & $0$ & $1$ & $1$& $1$ & $0$ & $0$ & $-3$
\end{tabular}
\end{equation} On the mirror manifolds, the
Greene-Plesser orbifold group acts as $x_i\rightarrow
\lambda_k^{g_{k,i}}x_i$ with weights \bqa
\mathbb{Z}_6:~g_1=(1,-1,0,0,0),~~~\mathbb{Z}_4:~g_2=(0,1,2,1,0)\eqa
where we denotes $\lambda_{1}^6=1,~\lambda_{2}^4=1$.

   In \cite{Feng}, we have obtained the period in subsystem as
   follows
  \bqa
  \begin{split}
  \eqlabel{solution}
  &\pi(u_1,u_2)=\frac{c}{2}B_{\{\hat{l_1},\hat{l_2}\}}(u_1,u_2;0,\frac{1}{2})
  \end{split}
  \eqa
  where $c$ are some normalization constants not determined by the
  differential operator.
According to \eqref{sub}, the off-shell superpotentials can be
obtained by integrating the $\pi$:
  \bqa
  \mathcal{T}_a^{\pm}(z_1,z_2,z_3)=\frac{1}{2\pi
  i}\int\pi(\hat{z})\frac{d\hat{z}}{\hat{z}},
\eqa with the appropriate integral constants\cite{Alim:2010za},
 the superpotentials can be chosen as  $\mathcal{W}^+=-\mathcal{W}^-$.

  Eventually, The superpotential are \bqa
\begin{split}
&\mathcal{W}^{\pm}(z_1,z_2,z_3,\hat{z})=\sum_{n_1,n_2,n_3}\frac{\mp cz_1^{\frac{1}{2}+n_1}z_2^{n_2}\hat{z}^{\frac{-1-2n_1}{2}}\Gamma(4n_1+\frac{5}{2})}{\Gamma(1+2n_2)\Gamma(1+n_2)\Gamma(\frac{3}{2}+n_1)\Gamma(2+2n_1)\Gamma(n_1-3n_2+\frac{3}{2})}\\
&\frac{\{(1-2n_1)
{}_2F_1(-\frac{1}{2}-n_1,-2n_1,\frac{1}{2}-n_1;\hat{z})+\hat{z}(1+2n_1){}_2F_1((\frac{1}{2}-n_1,-2n_1,\frac{3}{2}-n_1;\hat{z}))\}}{4\pi(-1+4n_1^2)}
\end{split}
\eqa

  For calculation of instanton corrections, one need to know mirror
  map. The fundamental period $\omega_0$ is solution of the Picard-Fuchs equation which we
  listed in \cite{Feng}.  The flat coordinates in
 A-model at large radius regime are related to the flat coordinates of B-model at large complex structure regime by mirror map$t_i=\frac{\omega_i}{\omega_0},~\omega_i:= D_i^{(1)}
 \omega_0(z,\rho)|_{\rho=0}$. The open-string mirror map are
 \bqa
 \begin{split}
 &q_1=z_1+40z_1^2+1876z_1^3+2z_1z_2-13z_1z_2^2+z_1z_2z_3+\mathcal{O}(z^4)\\
 &q_2=z_2-6z_2^2+63z_2^3+z_2z_3-9z_2^2z_3+\mathcal{O}(z^4)\\
 &q_3=z_3-z_3^2+z_3^3+\mathcal{O}(z^4)
 \end{split}
 \eqa
here $q_i=e^{2\pi i t_i}$ and we can obtain the inverse mirror map
as follows
 \bqa
\begin{split}
&z_1=q_1-40q_1^2+1324q_1^3-2q_1q_2+268q_1^2q_2+5q_1q_2^2+\mathcal{O}(q^4)\\
 &z_2=q_2+6q_2^2+9q_2^3-36q_1q_2-468q_1q_2^2+630q_1^2q_2+\mathcal{O}(q^4)\\
&z_3=q_3+q_3^2+q_3^3+\mathcal{O}(q^4)
 \end{split}
 \eqa

\begin{table}[!h]
\tabcolsep 0.01in
\def\temptablewidth{1.1\textwidth}
\begin{center}
\begin{tabular*}{480pt}{@{\extracolsep{\fill}}c|ccccc}
$d_3=0$&&&&&\\
$d_1/2\backslash d_2$ & 0 &1     &2     &3    &4 \\\hline
 1       &1       &  $\frac{-13}{16}$        & $\frac{2693}{1024}$           &$\frac{19517}{9}$        &$\frac{7703}{16384}$  \\
  3      & $\frac{1312}{243}$    &$\frac{68231}{1296}$                    & $\frac{-23305385}{82944}$        & $\frac{-3519745}{18}$                & $\frac{-1672979243}{3981312}$  \\
   5     & $\frac{63544513}{28350}$       & $\frac{135578197}{12960}$             & $\frac{346285919719}{5806080}$      & $\frac{-9330830923}{1944}$            & $\frac{-1608130586479}{92897280}$   \\
   7     &$\frac{172956753731}{1389150}$     & $\frac{5372183267179}{3175200}$        & $\frac{-21892937788889}{8128512}$    &$\frac{32917422417037}{136080}$             &$\frac{-282745996819463}{43352064}$   \\
   9     & $\frac{13409490308809711}{600112800}$ &$\frac{216480619417211431}{355622400}$ &$\frac{-19644707820777819881}{13655900160}$ &$\frac{-82475053081873279}{7620480}$ &$\frac{311040357663729110033}{93640458240}$
       \end{tabular*}
       {\rule{\temptablewidth}{0.5pt}}
\end{center}
       \end{table}
\begin{table}[!h]
\footnotesize
\def\temptablewidth{1.1\textwidth}
\begin{center}
\begin{tabular*}{\temptablewidth}{@{\extracolsep{\fill}}c|ccccc}
$d_3=1$&&&&&\\
$d_1/2\backslash d_2$ & 0         &1         &2     &3    &4
\\\hline
 1                &$\frac{-10 }{9}$      & $\frac{175}{144}$      & $\frac{-18923}{2304}$                         &$\frac{23077 }{147456}$                           &$\frac{-132267135 }{2097152}$  \\
  3              & $\frac{3380}{81}$         &$\frac{-82429}{432}$    & $\frac{48484423}{41472}$                       & $\frac{1376117443}{1327104}$               & $\frac{601661687053}{42467328}$  \\
   5              & $\frac{-1138840}{567}$     &$\frac{403544255}{18144}$        & $\frac{-534921106991 }{2903040}$         & $\frac{166478391791}{3440640}$           & $\frac{-20526886980289679 }{5945425920}$   \\
   7              &$\frac{3400299058}{42525}$      & $\frac{-835235546479}{272160}$    & $\frac{786036635335453}{60963840}$         &$\frac{4714357006892647}{650280960}$           &$\frac{-44180037787516945679}{62426972160}$   \\
   9              & $\frac{-121337433752293}{33339600}$ &$\frac{-99152104754391869}{152409600}$ &$\frac{15374804216369862857}{8534937600}$ &$\frac{-62693086142469434527}{6242697216}$  &$\frac{-923658211082431780070641}{3995326218240}$
       \end{tabular*}
       {\rule{\temptablewidth}{1pt}}
       \tabcolsep 0pt \caption{Disc invariants $n_{d_1,d_2,d_3}$ for the off-shell
superpotential $\mathcal{W}_1^{+}$ of the 3-fold
$\mathbb{P}_{1,1,1,3,6}[12]$}. \vspace*{-12pt}
\end{center}
       \end{table}

  Using the modified multi-cover
formula\cite{Aganagic:2000gs} for this case
 \bqa
\frac{\mathcal{W}^{\pm}(z(q))}{w_0(z(q))}=\frac{1}{(2\pi
i)^2}\sum_{k~
 odd}~\sum_{d_1 odd, d_{2,3}\geq
0}n_{d_1,d_2,d_3}^{\pm}\frac{q_1^{kd_1/2}q_2^{kd_2}q_3^{kd_3}}{k^2},
\eqa
 The superpotentials $\mathcal{W}^+$ give
 Ooguri-Vafa invariants $n_{d_1,d_2,d_3}$ for the normalization constants
  $c=1$, which are listed in Table. 2

\section{Summary}
  ~~~~In this paper, we make a further step of previous work \cite{Feng} and calculate the off-shell superpotential. By
open mirror symmetry, we also compute the Ooguri-Vafa invariants
from A-model expansion.

 The superpotential of Type II string theory are important in both
physics and mathematics. It also related to F-theory by open-closed
duality \cite{Mayr:2001xk,Jockers:2009ti,Alim:2009bx}. In type
II/F-theory compactification, the vacuum structure is determined by
the superpotentials, whose second derivative gives the chiral ring
structure. The quantum cohomology ring structure comes from the
world-sheet instanton corrections and space-time instanton
corrections\cite{Lerche:2002ck,Lerche:2002yw}. In fact, the more
general vacuum structure of type II/F-theory/heterotic theory
compactification can be tackled in Hodge variance approach.

  In next work, we will study D-brane in general case. We also try to calculate the D-brane
superpotential
  with the method of
  $A_{\infty}$ structure of the derived category $D_{\text{coh}}(X)$ and path algebras
  of quivers.

\section*{Acknowledgments}
~~~~ The work is supported by the NSFC (11075204) and President Fund
of GUCAS (Y05101CY00).

\end{document}